\documentclass[aps]{revtex4}
\usepackage{graphicx}
\usepackage{amsmath}
\usepackage{amssymb}
\newcommand{\bb}{\begin{eqnarray}}
\newcommand{\ee}{\end{eqnarray}}
\usepackage{subfig}
\usepackage{graphicx}
\begin{document}

\title{\bf Tunneling across dilaton coupled black holes in anti de Sitter spacetime}
\author{Tanwi Ghosh\footnote{E-mail: tanwi.ghosh@yahoo.co.in} and Soumitra SenGupta\footnote{E-mail: tpssg@iacs.res.in}}
\affiliation{Department of Physics,Ranaghat College,\\ District- Nadia,Pin-741201\\
 and Department of Theoretical Physics , Indian Association for the
Cultivation of Science,\\
Jadavpur, Calcutta - 700 032, India}
\vskip 5cm

\begin{abstract}
Considering generalised action for dilaton coupled Maxwell-Einstein theory in four dimensions ,Gao 
and Zhang obtained black holes solutions for asymptotically anti de Sitter ( Ads ) and de Sitter ( ds ) spacetimes. 
We study the Hawking radiation in  Parikh-Wilczek's tunneling formalism as well as using 
Bogoliubov transformations. We compare the expression of the Hawking temperature obtained from these two different approaches.
Stability and the extremality conditions for such black holes are discussed. The exact dependences of the Hawking temperature and flux
on the dilaton coupling parameter are determined. It is shown that the Hawking flux increases with the dilaton coupling
parameter. 
Finally we show that the expression for the Hawking flux obtained using Bogoliubov transformation matches exactly with flux
calculated via chiral gauge and gravitational anomalies. This establishes a correspondence among all these different approaches
of estimating Hawking radiation from these class of black holes.  
\end{abstract}
\maketitle

{{\Large {\bf Introduction}}}\\

Hawking radiation from black holes \cite{hawk,hawk2} is known to be a quantum phenomena.
Parikh and Wilczek viewed this as a quantum tunneling of particles through black hole horizon\cite{pw1,
pw2,pw3}.Energy conservation as well as particle's self gravitation effects are taken into account in this
approach. Introducing Painleve coordinate transformation,
which is well behaved across the horizon, Parikh and Wilczek found a generic relation between the entropy change and the 
tunneling rate which is valid for a large class of black holes. In an alternative approach, Bogoliubov transformation 
is used in two different coordinates (such as Kruskal coordinates and asymptotic coordinates ) to compute the coefficients of the
energy modes. In this approach, neglecting the back scattering effect and normalising Bogoliubov coefficients, the
leading thermal characteristic of Hawking radiation has been derived. In yet another approach the Hawking flux has been 
evaluated via chiral gauge and gravitational anomalies from a covariant boundary condition. Equivalence of these
approaches in respect to thermodynamic behaviour of black holes has been an important area of study. 
There have been several works in recent times to explore Hawking radiation for different kinds of black holes using these approches\cite{he,aj,aj1,ar,va,va1,va2,ra1,ra2,ra3,ra4,debraj,al,zh1,zh2,zh3,ji,fang,zh4,cheng,quing,arzano,wu,hu,shu,ji1,ji2,de,yu,ran,
gu,heng,ren,ya1,ma,liu,shu1,shu2,iso}. \\
Meanwhile Gao and Zhang found the dilaton coupled de Sitter and anti de Sitter black hole solutions using appropriate 
Liouville type potential for the dilaton field.  
Dilaton coupled black hole has it's origin rooted into string theory which in the low energy effective field theory
appears naturally with a dilaton-electromagnetic coupling.  Anti de Sitter space time has received interest in the context of Ads/Cft correspondence \cite{rei,mal,stro} and also in brane world scenarios of Randall-Sundram model\cite{ran1}. Our objective here is to study Hawking radiation for the class of black holes using quantum tunneling procedure, computing Bogoliubov coefficients as well as estimating gauge and gravitational anomalies to compare the leading thermal behaviour of radiation spectrum. Our results establish the equivalence of these descriptions of Hawking radiation. The dilaton coupling parameter is shown to enhance the Hawking flux significantly. \\ 

{{\Large {\bf The Model }}}.\\

The action describing dilaton black hole in both de Sitter and anti de Sitter space time can be expressed as follows\cite{gao}:\\
\begin{eqnarray}
S=\int d^4x \sqrt{-g}[R-2\partial_{\mu}\varphi\partial^{\mu}\varphi-e^{-2\alpha\varphi}F^2
-V(\varphi)]
\end{eqnarray}
where $V(\varphi)=\frac{-2\lambda}{3(1+\alpha^2)^2}[\alpha^2(3\alpha^2-1)e^{\frac{-2\varphi}{\alpha}}
+(3-\alpha^2)e^{2\alpha\varphi}+8\alpha^2e^{\varphi\alpha-\frac{\varphi}{\alpha}}]$ is
the dilaton potential for  anti de Sitter black holes, $\varphi$ is the dilaton field. $\alpha$
represents coupling parameter of the scalar field with Maxwell field $F_{\mu\nu}$ and $\lambda$ 
is the cosmological constant.\\

The interpretation of the parameter $\lambda$ as the cosmological constant has been discussed in \cite{gao} 
where starting from a dilaton de Sitter metric in cosmic coordinate system a scale factor $a=e^{Ht}$ was obtained 
with H as the Hubble constant. Replacing
$H^2=\frac{\lambda}{3}$,dilaton anti de Sitter/de-Sitter  black hole solution has been produced.\\

{{\Large {\bf Hawking Radiation By Uncharged Particle Tunneling From dilaton-anti De Sitter Black hole }}}.\\

Considering the general action describing dilaton field coupled to electromagnetic field in presence of three Liouville 
type potential, Gao and Zhang obtained dilaton black hole solutions in 
both de Sitter and anti de Sitter spacetime.
Here we study the phenomena of particle tunneling across the dilaton anti de Sitter black hole which is described by the 
metric\cite{gao}, 
\begin{eqnarray}
ds^2&=&-f(r)dt_s^2+g(r)dr^2+h(r)[d\theta^2+sin^2\theta d\phi^2]\\
&&=-[(1-\frac{r_+}{r})(1-\frac{r_-}{r})^{\frac{(1-\alpha^2)}{(1+\alpha^2)}}-\frac{\lambda}{3}r^2 \nonumber(1-\frac{r_-}{r})^{\frac{2\alpha^2}{(1+\alpha^2)}}]dt_s^2\\
&&+[(1-\frac{r_+}{r})(1-\frac{r_-}{r})^{\frac{(1-\alpha^2)}{(1+\alpha^2)}}
\nonumber-\frac{\lambda}{3}r^2 (1-\frac{r_-}{r})^{\frac{2\alpha^2}{(1+\alpha^2)}}]^{-1}dr^2\\
&&+r^2(1-\frac{r_-}{r})^{\frac{2\alpha^2}{(1+\alpha^2)}}d\Omega
\end{eqnarray}
The event horizons of the black hole $r_+$ ,$r_-$, elctromagnetic charge Q 
and asymptotic value of dilaton field $\varphi_0$ are related 
through $e^{2\alpha\varphi_0}=\frac{r_+r_-}{(1+\alpha^2)Q^2}$ for arbitrary $\alpha$.
For $\alpha =1$, the expressions of f(r) and h(r) for both dilaton anti de Sitter and de Sitter black hole having mass M and dilaton charge $D=\frac{Q^2 e^{2\varphi_0}}{2M}$ respectively take the following form
\begin{eqnarray}
f(r)=\frac{1}{g(r)}=1-\frac{2M}{r}-\frac{\lambda}{3}r(r-2D)
\end{eqnarray}
\begin{eqnarray}
f(r)=\frac{1}{g(r)}=1-\frac{2M}{r}-H^2r(r-2D)
\end{eqnarray}
and
\begin{eqnarray}
h(r)= r(r-2D)
\end{eqnarray}

Consider a tunneling particle with energy $\omega$ and charge e across such a  dilaton anti de Sitter  black hole such that the total energy of the system comprising of the black hole and the particle is conserved.\\

As the  anti de Sitter Painleve coordinates have the same geometry as global anti de Sitter metric of constant time slices, 
we choose the following Painleve-type coordinate transformation\cite{ma}.
\begin{eqnarray}
dt_s=dt-\sqrt{\frac{g(r)-1}{f(r)}}dr
\end{eqnarray}
After this transformation line element (2) can be written as,
\begin{eqnarray}
ds^2=-f(r)dt^2+2\sqrt{f(r)}\sqrt{g(r)-1}dtdr+dr^2+h(r)(d\theta^2+sin^2{\theta}d\phi^2)
\end{eqnarray}
The advantage of Painleve coordinate transformation is that none of the components of the metric
or the inverse metric diverge.
The radial velocity of the outgoing particle in such a coordinate is obtained as,
\begin{eqnarray}
\dot{r}=\sqrt{f(r).g(r)}(1-\sqrt{1-\frac{1}{g(r)}})
\end{eqnarray}
During particle tunneling black hole mass M  should be replaced by $M-\omega$.
Following Parikh-Wilczek tunneling formalism \cite{pw1,pw2,pw3,ma} ,the imaginary part of the action 
for an outgoing particle crossing the horizon of the black holes can be writen as,
\begin{eqnarray}
Im I=Im\int_{r_h(M)}^{r_h(M-\omega)}\int_0^{p_r}dp'_r dr
\end{eqnarray}
where within the integration we have used $p'_r$ for canonical momentum conjugate to radial variable r.
Considering particle's self gravitational effect \cite{kraus1,kraus2} and using Hamilton's equations 
we calculate the radial velocity in the dragged frame. In this frame, by an appropriate transformation, the black hole
event horizon can be made to coincide with infinite redshift surface so that WKB approximation is valid. Thus radial
velocity and time derivative of vector potential $\dot{A_t}$ in the dragged frame becomes,
\begin{eqnarray}
\dot{r}=\frac{dH}{dp_r}=\frac{d(M-\omega)}{dp'_r}
\end{eqnarray}
and
\begin{eqnarray}
\dot{A_t}=\frac{dH}{dp'_{A_t}}
\end{eqnarray}
where $p_r$ and H are the canonical momentum conjugate to radial variable r  and energy of the black hole respectively.
Substituting equation (11) and (12)in (10),we get
\begin{eqnarray}
Im I=Im\int_{r_h(M),Q}^{r_h(M-\omega),Q-e}\int_0^{\omega}\frac{d(M-\omega',Q-e')dr}{\dot{r}}
=Im\int_{r_h(M),Q}^{r_h(M-\omega),Q-e}\int_0^{\omega}\frac{(dM'-\frac{Q'dQ'e^{2\alpha\varphi_0}}{r})dr}{\dot{r}}
\end{eqnarray}
where $M'=M-\omega'$ ,$Q'=Q-e$,$r_h(M)$ and $r_h(M-\omega)$ are the locations of the horizon before and after the emission of particle.
Using the expression for $\dot{r}$ from equation (9), $Im I$ takes the form,
\begin{eqnarray}
Im I =Im\int_0^{\omega,Q}\int_{r_h(M-\omega,Q-e)}^{r_h(M),Q}\frac{dr d\omega'}{\sqrt{f(r)g(r)}(1-\sqrt{1-\frac{1}{g(r)}})}
\end{eqnarray}
Now using WKB approximation the tunneling rate of the particle can be expressed as
\begin{eqnarray}
\Gamma=\Gamma_0 e^{-2Im{I}}
\end{eqnarray}
where $\Gamma_0 $ is the normalisation constant.
From equation (2),it follows that for four-dimensional dilaton anti de Sitter black hole the function $g(r)$ has a singularity 
at $r=r_h$.Thus $g(r)$ can be written as,
\begin{eqnarray}
g(r)=\frac{C(r)}{(r-r_h)}
\end{eqnarray}
where the function $C(r)$ is regular on the horizon. Now substituting (16) in (13) and simplifying, we have the following equation,
\begin{eqnarray}
Im{I}=Im\int_0^{\omega}\int_{r_h(M-\omega,Q-e)}^{r_h(M,Q)}dr 
(dM'-\frac{Q'dQ'e^{2\alpha\varphi_0}}{r_h})\frac{1+\sqrt{1-\frac{r-r_h}{C(r)}}}
{\sqrt{f(r)g(r)}\frac{r-r_h}{C(r)}}
\end{eqnarray}
We deform the integral around the simple pole at $r = r_h$ and obtain, 
\begin{eqnarray}
Im{I}=2\pi\int_0^{\omega}\frac{C(r_h)}{\sqrt{f(r_h)g(r_h)}}(dM'-\frac{Q'dQ'e^{2\alpha\varphi_0}}{r_h})
\end{eqnarray}
For such black holes,f(r)and $g^{-1}(r)$ are regular near the horizon. Thus expanding  f(r) and $g^{-1}(r)$ near the horizon,we get
\begin{eqnarray}
f(r)=f'(r_h)(r-r_h)+O(r-r_h)^2
\end{eqnarray}
\begin{eqnarray}
g^{-1}(r)=g'^{-1}(r_h)(r-r_h)+O(r-r_h)^2
\end{eqnarray}
Again from equation (16) we readily obtain,
\begin{eqnarray}
g'^{-1}(r_h)=\frac{1}{C(r_h)}
\end{eqnarray}
Using equations (18), (19) and (20),we want to obtain the expression of Im I from equation (17) which can be written 
as,
\begin{eqnarray} 
Im{I}=\frac{1}{2}\int \frac{(dM'-\frac{Q'dQ'e^{2\alpha\varphi_0}}{r_h})}{\frac{\kappa(r_h)}{2\pi}}
\end{eqnarray}
where $\kappa(r_h)=\frac{\partial_r f(r)}{2}|_{r=r_h}$.
To explore the thermal character of such  tunneling, we integrate equation (22) and obtain, 
\begin{eqnarray}
Im I = \frac{\pi(\omega-\frac{Qe^{2\alpha\varphi_0}}{r_h})}{\kappa(r_h)}
\end{eqnarray}
Sustituting (22) in (14) we obtain the expression of the tunneling rate as,
\begin{eqnarray}
\Gamma=\Gamma_0 e^{\frac{-2\pi(\omega-\frac{Qe^{2\alpha\varphi_0}}{r_h})}{\kappa(r_h)}}
\end{eqnarray}
According to Boltzmann distribution for the emission of a particle having energy $\omega$ and charge e the emission rate of such a black hole can be represented by,
\begin{eqnarray}
\Gamma= \Gamma_0 e^{\frac{-(\omega-\frac{Qe^{2\alpha\varphi_0}}{r_h})}{T_h}}
\end{eqnarray}
where $T_h$ is the black hole temperature. 
Compairing equation (23) with (25),one can obtain the expression of temperature in terms of surface gravity
$\kappa(r_h)$ as,
\begin{eqnarray}
T_h=\frac{\kappa(r_h)}{2\pi}=\frac{\partial_r f(r)}{4\pi}|_{r=r_h}
\end{eqnarray}
Finally particle's tunneling probability P in this approach can be expressed as,
\begin{eqnarray}
P=exp{(-\frac{(\omega-\frac{Qe^{2\alpha\varphi_0}}{r_h})}{T_h})}
\end{eqnarray}
Using (26),the general expression of temperature for dilaton anti de Sitter black hole having 
arbitrary coupling $\alpha$ can be expressed as,
\begin{eqnarray}
T_h=\frac{1}{4\pi}[r_+ r_h^{\frac{-(3+\alpha^2)}{(\alpha^2+1)}}(r_h-\frac{(\alpha^2+1)Q^2e^{2\alpha\varphi_0}}{r_+})
^{\frac{(1-\alpha^2)}{(1+\alpha^2)}}+\frac{\lambda Q^2e^{2\alpha\varphi_0}}{3r_+r_h^
{\frac{(\alpha^2-1)}{(\alpha^2+1)}}}{(1-3\alpha^2)}\\
\nonumber(r_h-\frac{(\alpha^2+1)Q^2e^{2\alpha\varphi_0}}{r_+})^{\frac{(\alpha^2-1)}{(1+\alpha^2)}}-\frac{\lambda}{3}2r_h^
{\frac{(1-\alpha^2)}{(1+\alpha^2)}}(r_h-\frac{(\alpha^2+1)Q^2e^{2\alpha\varphi_0}}{r_+})^
{\frac{2\alpha^2}{(\alpha^2+1)}}]
\end{eqnarray}
where the location of horizon position can be determined from the following ,
\begin{eqnarray}
(1-\frac{r_+}{r_h})(1-\frac{r_-}{r_h})^{\frac{(1-\alpha^2)}{(1+\alpha^2)}}-\frac{\lambda}{3}
r_h^2(1-\frac{r_-}{r_h})^{\frac{2\alpha^2}{(\alpha^2+1)}}=0
\end{eqnarray}
Considering dilaton de Sitter black holes for $\alpha=1$ having mass
M and dilaton charge D and using the transformation \cite{gao},$r=D+\sqrt{x^2+D^2}$,the expression for $T_h$ is given by, 
\begin{eqnarray} 
T_h =\frac{1}{4\pi}[\frac{2M r_h}{\sqrt{r_h^2+D^2}(D+\sqrt{r_h^2+D^2})^2}-2r_h H^2]
\end{eqnarray}
where $r_h$ can be determined from 
\begin{eqnarray}
1-\frac{2M}{(D+\sqrt{r_h^2+D^2})}-H^2r_h^2=0
\end{eqnarray}
Near the extremal case cosmological horizon lies very near to black hole event horizon. In the extremal case for M=D,
black hole event horizon disappears and only one cosmic horizon will survive.
For non-extremal black holes we have plotted the expression of temperature of dilaton anti de Sitter black hole
considering $\alpha=1$,M=1and $\varphi_0=0$ with respect to cosmological constant $\lambda$ and charge Q in Fig1.This 
figure reveals the fact that dilaton black hole temperature increases as $-\lambda$ increases as well as
charge decreases. \\
\begin{figure}[h] 
\includegraphics[width=3.330in,height=2.20in]{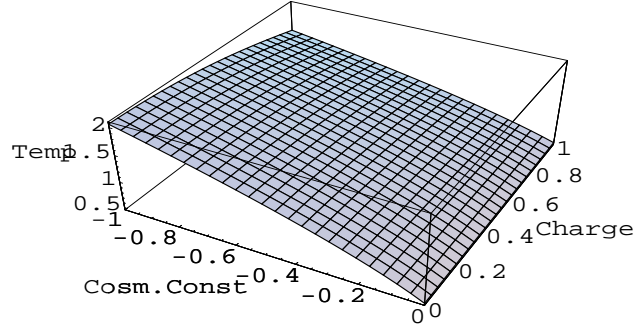}
\caption{Graph of temterature $T_h$ (Temp)versus cosmological constant $\lambda$ (Cosm.Const)and charge Q} \label{radphot}
\end{figure}
In this context the location of the radius of the horizon for such kind of black hole
can be determined from the equation,
\begin{eqnarray}
r_h^3-\frac{Q^2 e^{2\varphi_0}}{M}r_h^2-\frac{3}{\lambda}r_h+\frac{6M}{\lambda}=0
\end{eqnarray}
We have also plotted the expression of horizon radius with respect to cosmological constant $\lambda$
and charge Q in Fig2. Substituting M=1 and $\varphi_0=0$ we find that horizon radius decreases slowly as 
$-\lambda$ increases.\\ 

\begin{figure}[h] 
\includegraphics[width=3.330in,height=2.20in]{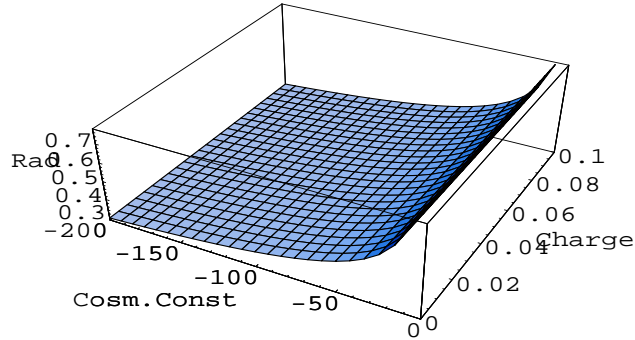}
\caption{Graph of horizon $r_h$ (Rad)versus cosmological constant $\lambda$ (Cosm.Const)and charge Q} \label{radphot}
\end{figure}

{{\Large {\bf Hawking Radiation From Dilatonic Anti de Sitter Black hole Using Bogoliubov Transformation:}}}\\

We would now like to describe Hawking radiation from a different approach,i.e,using Bogoliubov transformations of creation and 
annihilation operators between two basis. We use mode expansions of the fields in two different coordinate systems such as 
asymptotic coordinates and Kruskal coordinates. We apply normalisation condition between the coefficients of positive
and negative energy modes and finally obtain the tunneling probability measured by an observer siting outside the horizon.
Let us introduce tortoise coordinate $r_*$ for black holes represented by equations (4) and (6) as,
\begin{eqnarray}
r_{*}=\int_{r_h}^r\frac{dr'}{f(r')}
=\frac{A}{2a}[ln(ar^2+br+n)-ln(ar_h^2+br_h+n)]+\frac{C}{g}[ln(\frac{gr}{p}+1)-ln(\frac{gr_h}{p}+1)]+\\\nonumber
\frac{(B-\frac{bA}{2a})}{\sqrt{a(n-\frac{b^2}{4a})}}[tan^{-1}(\frac{(r+\frac{b}{2a})\sqrt{a}}{\sqrt{(n-\frac{b^2}{4a})}})
-tan^{-1}(\frac{(r_h+\frac{b}{2a})\sqrt{a}}{\sqrt{(n-\frac{b^2}{4a})}})]
\end{eqnarray}
where $a=\frac{-\lambda}{3}$,$b=\frac{\lambda p}{3}+\frac{2D\lambda}{3}$,$n=\frac{-2M}{p}$,$g=1$,$A=\frac{\lambda C}{3}$,
$B=[1-(\frac{2\lambda p}{3}+\frac{2D\lambda}{3})C]$,$C=\frac{p}{[p(\frac{2\lambda p}{3}+\frac{2D\lambda}{3})+\frac{2M}{p}]}$ and 
p can be determined from
\begin{eqnarray}
\frac{\lambda}{3}p^3+\frac{2D\lambda}{3}p^2-p-2M=0
\end{eqnarray}
Introducing null coordinates u and v as,
\begin{eqnarray}
u=t-r_*\\\nonumber
v=t+r_*
\end{eqnarray}
the above dilaton anti de Sitter black hole metric (given by equation (4) and (6))  take the form as,
\begin{eqnarray}
ds^2=-f(r)dudv +h(r)d\Omega^2
\end{eqnarray}
Using Kruskal coordinates U and V as,
\begin{eqnarray}
U=-exp(-2\pi T_h u)\\\nonumber
V=exp(2\pi T_h v)
\end{eqnarray}
where the Hawking temperature $T_h$ is defined by equation (24),we will get 
\begin{eqnarray}
ds^2=-\frac{f(r)e^{-4\pi T_h r^*}}{(2\pi T_h)^2}dU dV +h(r)d\Omega^2
\end{eqnarray}
Substituting the expressions of temperature ,$r_*$ and $f(r)$,the black hole metric reduces in Kruskal coordinates as follows,
\begin{eqnarray}
\nonumber ds^2=-4\frac{[1-\frac{2M}{r}-\frac{\lambda}{3}r(r-2D)]}{[\frac{2M}{r_h^2}-\frac{\lambda}{3}(2r_h-2D)]^2}\\
exp[-(\frac{2M}{r_h^2}-\frac{\lambda}{3}(2r_h-2D))(\frac{A}{2a}[ln(ar^2+br+n)-ln(ar_h^2+br_h+n)]+C[ln(\frac{r}{p}+1)-
ln(\frac{r_h}{p}+1)]+\nonumber\\
\frac{(B-\frac{bA}{2a})}{\sqrt{a(n-\frac{b^2}{4a})}}[tan^{-1}(\frac{(r+\frac{b}{2a})\sqrt{a}}{\sqrt{(n-\frac{b^2}{4a})}})
-tan^{-1}(\frac{(r_h+\frac{b}{2a})\sqrt{a}}{\sqrt{(n-\frac{b^2}{4a})}})])]dUdV +h(r)d\Omega^2
\end{eqnarray}
To describe quantum fields in black hole background,one requires canonical choices of natural vacuum states.
Let us define different energy modes corresponding to fields in the vacuum.
Following \cite{ruh,he},we can define the different kinds of natural modes having frequency $\omega$ and charge e
for this kind of black holes as follows,
\begin{eqnarray}
\phi_{+,\omega}=exp(-i\omega u)=\theta(-U)(-U)^{\frac{i(\omega-\frac{eQe^{2\alpha\varphi_0}}{r_h})}{2\pi T_h}}(U<0)\\
\phi_{-,\omega}=exp(i\omega u)=\theta(U)(U)^{\frac{-i(\omega-\frac{eQe^{2\alpha\varphi_0}}{r_h})}{2\pi T_h}}(U>0)
\end{eqnarray}
and
\begin{eqnarray}
\phi_{1,\omega}=\theta(-U)(-U)^{\frac{i(\omega-\frac{eQe^{2\alpha\varphi_0}}{r_h})}{2\pi T_h}}+
C_1\theta(U)(U)^{\frac{i(\omega-\frac{eQe^{2\alpha\varphi_0}}{r_h})}{2\pi T_h}}\\
\phi_{2,\omega}=\theta(U)(U)^{\frac{-i(\omega-{eQe^{-\frac2\alpha\varphi_0}}{r_h})}{2\pi T_h}}+
C_2\theta(-U)(-U)^{\frac{-i(\omega-\frac{eQe^{2\alpha\varphi_0}}{r_h})}{2\pi T_h}}
\end{eqnarray}
Using boundary conditions one can find out,
\begin{eqnarray}
C_1=C_2\equiv C=exp[-\frac{(\omega-\frac{eQe^{2\alpha\varphi_0}}{r_h})}{T_h}]
\end{eqnarray}
Bogoliubov coefficients $\alpha$ and $\beta$ are defined as follows,
\begin{eqnarray}
\alpha=(\phi_{1,\omega},\phi_{+,\omega})\\
\beta=(\phi_{1,\omega},\phi_{-,\omega}^*)
\end{eqnarray}
The normalisatin conditions between $\alpha$ and $\beta$ gives the average occupation number which can be 
simplified to the Fermi distribution function as below.
\begin{eqnarray}
n_{\omega}=\frac{1}{[exp(\frac{(\omega-\frac{eQe^{2\alpha\varphi_0}}{r_h})}{T_h})+1]}
\end{eqnarray}
The ratio of Bogoliubov coefficients thus gives the expression of tunneling probability P as,
\begin{eqnarray}
P= |C|^2=exp{(\frac{-(\omega-\frac{eQe^{2\alpha\varphi_0}}{r_h})}{T_h})}
\end{eqnarray}
Compairing (27) and (48) we find the same expression of tunneling probability with Hawking temperature $T_h$ as was
obtained using tunneling formalism.
The expression of Hawking flux can easily be obtained by integrating equation (47) for all $\omega$ as,
\begin{eqnarray}
Flux = \frac{{\hbar}^2\kappa^2}{48\pi}+\frac{e^2Q^2e^{4\alpha\varphi_0}}{4\pi r_h^2}
     =\frac{\pi}{12\beta^2}+\frac{e^2Q^2e^{4\alpha\varphi_0}}{4\pi r_h^2}
\end{eqnarray}\\
where $\frac{1}{T_h}=\beta$

{{\Large {\bf Hawking Radiation Flux For Dilatonic Anti de Sitter Black hole Using Chiral Gauge and Gravitational
Anomalies:}}}\\

In addressing Hawking radiation through yet another alternative 
approach,we now derive Hawking flux using chiral gauge and gravitational anomalies.
In this approach,the expression of the flux of energy-momentum tensor\cite{ra1}obtained in the asymptotic limit for dilaton anti de sitter blackhole can be written as,
\begin{eqnarray}
T^{r}_{t}(r\rightarrow\infty)=\frac{e^2}{4\pi}A_t^2(r_h)+\frac{1}{192\pi}f'^2(r_h)
\end{eqnarray}
where the form of vector potential
is $A_t=\frac{Qe^{2\alpha\varphi_0}}{r_h}$. 
Substituting the expression of vector potential $A_t$ and $f'^2(r_h)$ in (50) for our black hole we get,
\begin{eqnarray}
T^{r}_{t}(r\rightarrow\infty)= \frac{e^2Q^2e^{4\alpha\varphi_0}}{4\pi r_h^2}+\frac{\pi}{12\beta^2}
\end{eqnarray}
Equation (51) can be compared with the fluxes as obtained from the blackbody radiation.
Using Planck distribution function for fermions \cite{iso} (avoiding superradiance)as,
\begin{eqnarray}
J^{\pm}(\omega)=\frac{1}{[exp(\frac{\omega\pm\frac{eQe^{2\alpha\varphi_0}}{r_h}}{T_h})+1]}
\end{eqnarray}
the expression of flux of energy-momentum tensor given by equation (51) can be rederived as,
\begin{eqnarray}
T^{r}_{t}=\int_{0}^{\infty}\frac{\omega}{2\pi}d\omega[J^{-}(\omega)+J^{+}(\omega)]
=\int_{0}^{\infty}\frac{\omega}{2\pi}d\omega 
[\frac{1}{(exp(\frac{\omega+\frac{eQe^{2\alpha\varphi_0}}{r_h}}{T_h})+1)}+
\frac{1}{(exp(\frac{\omega-\frac{eQe^{2\alpha\varphi_0}}{r_h}}{T_h})+1)}]
=\frac{e^2Q^2e^{4\alpha\varphi_0}}{4\pi r_h^2}+\frac{\pi}{12\beta^2}
\end{eqnarray}\\
Thus equation (51) derived from chiral anomaly cancellation gives identical result as obtained from black body
radiation. Our conclusion is black hole thermal flux is able to cancel the anomaly.\\

We conclude our analysis with the following observations.
If we  consider Euclidean dilaton-Ads black hole in the asymptotic limit,
\begin{eqnarray}
ds^2=[1-\frac{(1-\alpha^2)Q_1^2e^{2\alpha\varphi_0}}{r^2}-\frac{\lambda}{3}r^2]d\tau^2
+[1-\frac{(1-\alpha^2)Q_1^2e^{2\alpha\varphi_0}}{r^2}-\frac{\lambda}{3}r^2]^{-1}dr^2
+r^2d\Omega
\end{eqnarray}
where $Q_1^2=iQ$ is an analytic continuation of the electric charge. Then following the procedure shown in
\cite{hor} the expression of area A and the integral of extrinsic curvature $K$ respectively becomes,
\begin{eqnarray}
A=4\pi r^2\beta [1-\frac{(1-\alpha^2)Q_1^2e^{2\alpha\varphi_0}}{r^2}-\frac{\lambda}{3}r^2]^{\frac{1}{2}}
\end{eqnarray}
and
\begin{eqnarray}
\int K = 8\pi r\beta-4\pi\beta(1-\alpha^2)\frac{Q_1^2e^{2\alpha\varphi_0}}{r}-4\pi r^3\beta\lambda
\end{eqnarray}
Thus the expression of energy E is obtained as:
\begin{eqnarray}
E=\frac{-1}{8G l}[4\beta(1-\alpha^2)Q_1^2 e^{2\alpha\varphi_0}]
\end{eqnarray}
Here $l^2=\frac{-3}{\lambda}$. 
The above expression clearly brings out the dependence of energy on the dilaton coupling parameter $\alpha$.
The inverse temperature $\beta=\frac{\pi l}{r_h}$ has been 
calculated for $r_h>>l$. Substituting the expression of $\beta$ and using the expression
\begin{eqnarray}
(1-\alpha^2)Q_1^2e^{2\alpha\varphi_0}=\frac{r_h^4+r_h^2l^2}{l^2}
\end{eqnarray}
the energy can be written as,
\begin{eqnarray}
E\approx -\frac{\pi^4 l^5}{2G\beta^3}
\end{eqnarray}
The above expression is similar to that for Ressiner-Nordstrom black hole\cite{hor}.\\
It may further be mentioned that through Ads/CFT correspondence our analysis for such  black holes should be related to a 
dual gauge theory on $S^2 \otimes S^1$ with $\beta$ (the inverse temperature) being the period of $S^1$ and $l$ is the radius of $S^2$.
A negative energy proportional to $\beta^{-3}$ ( just as in equ.59 ) is therefore expected to arise from the Casimir effect in the 
quantum field theory calculation on  $S^2 \otimes S^1$.\\ 
 
{{\Large {\bf Conclusion:}}}\\

We have considered three different approaches to address Hawking radiation across dilaton anti de Sitter black holes.
In deriving our results Painleve coordinate transformation is used in Parikh -Wilczek's tunneling method to eliminate the 
coordinate singularity. In analogy to the flat black holes, Ads Painleve coordinate matches with the constant time slices 
of a global Ads black hole metric. In Bogoliubov transformation method to address Hawking radiation,tortoise coordinate 
has been introduced.After normalising Bogoliubov coefficients corresponding to different energy modes leading 
thermal behaviour of radiation spectra has been obtained. We have plotted horizon radius and also the temperature 
of such black holes to examine their dependence on cosmological constant as well as on charge. We also compare the 
expression of Hawking temperature in these two different methods to establish an equivalence between these two apparently 
distinct description of Hawking radiation. Finally we estimate the Hawking flux through the chiral gauge and gravitational 
anomalies which exactly matches with the expression of flux calculated via Bogoliubov transformation approach.
This explicitly establishes equivalence among different approaches that are followed to explain Hawking radiation.
The Hawking flux is shown to increase exponentially with the dilaton coupling parameter $\alpha$.
The location of the horizon as well as the Hawking temperature are also  shown to have non-trivial dependence on the dilaton coupling parameter.  
Deriving the expression of the specific heat we have shown that a two parameter family namely the cosmological constant and the dilaton coupling determine the stability of these class of black holes. Finally obtaining the expressions of energy and extremality condition we conclude with some comments about the possible Ads/CFT correspondence in the context of such dilaton anti de Sitter black holes.\\

{{\Large {\bf Acknowledgement:}}}\\

T.G wishes to thank DST(Govt of India) for financial support through Fast Track Proposal Ref No:SR/FTP/PS-16,2008.\\

\end{document}